\documentclass[10pt]{article}
\usepackage{graphicx}
\usepackage{lineno}

\def\Title#1{\begin{center} {\Large #1 } \end{center}}
\def\Author#1{\begin{center}{ \sc #1} \end{center}}
\def\Address#1{\begin{center}{ \it #1} \end{center}}

\newcommand\pubblock{\rightline{\begin{tabular}{l} Proceedings of the Second Annual LHCP\\ \pubnumber\\
         \pubdate  \end{tabular}}}

\newenvironment{Abstract}{\begin{quotation} \begin{center} 
             \large ABSTRACT \end{center}\bigskip 
      \begin{center}\begin{large}}{\end{large}\end{center} \end{quotation}}

\newenvironment{Presented}{\begin{quotation} \begin{center} 
             PRESENTED AT\end{center}\bigskip 
      \begin{center}\begin{large}}{\end{large}\end{center} \end{quotation}}

\def\beq{\begin{equation}}
\def\eeq#1{\label{#1}\end{equation}}
\def\eeqn{\end{equation}}

\def\beqa{\begin{eqnarray}}
\def\eeqa#1{\label{#1}\end{eqnarray}}
\def\eeqan{\end{eqnarray}}

\let\bar=\overbar

\def\Dslash{\not{\hbox{\kern-4pt $D$}}}
\def\dslash{\not{\hbox{\kern-2pt $\del$}}}

\def\msb{{\bar{\ssstyle M \kern -1pt S}}}

\usepackage{color}

\newcommand{\Ppsi}{$\rm{\psi(2S)}$}
\newcommand{\Jpsi}{$\rm{J/\psi}$}
\newcommand{\pt}{$p_{\rm{T}}$}

\newcommand\pubnumber{ }

\newcommand\pubdate{\today}

\def\affiliation{
On behalf of the ALICE Collaboration, \\
Faculty of Nuclear Sciences and Physical Engineering\\
Czech Technical University, Prague, Czech Republic}

\begin{document}

\large
\begin{titlepage}
\pubblock

\vfill
\Title{  Charmonium photoproduction in ultra-peripheral p-Pb and Pb-Pb collisions at the LHC with the ALICE experiment  }
\vfill

\Author{ Michal Broz  }
\Address{\affiliation}
\vfill
\begin{Abstract}

Vector mesons are copiously produced in ultra-peripheral collisions.

In these collisions, the impact parameter is larger than the sum of the radii of two projectiles, implying that electromagnetic processes become dominant.
The cross section for this process is sensitive to the gluon distribution and can therefore probe nuclear gluon shadowing (Pb-Pb) and the gluon structure function in the nucleon (p-Pb).

The ALICE Collaboration has performed the first measurement of the coherent J/$\psi$ and $\psi$(2S) photoproduction cross section in Pb-Pb collisions and that for exclusive J/$\psi$ photoproduction off protons in ultra-peripheral proton-lead collisions at the LHC. The results are compared to STARLIGHT and to QCD based models.

\end{Abstract}
\vfill

\begin{Presented}
The Second Annual Conference\\
 on Large Hadron Collider Physics \\
Columbia University, New York, U.S.A \\ 
June 2-7, 2014
\end{Presented}
\vfill
\end{titlepage}
\def\thefootnote{\fnsymbol{footnote}}
\setcounter{footnote}{0}
%

\normalsize 

\section{Introduction}
Collisions between heavy-ions and a proton/ion at the LHC can be used to study particle production in photonuclear interactions \cite{Ref:Baltz}. These interactions may occur in ultra-peripheral collisions
(UPC), where the impact parameter is larger than the sum of the radii of the two projectiles and here the hadronic interactions are strongly suppressed. The photon flux is proportional to the square of the nucleus charge, so the photon flux in lead beams is enhanced by nearly four orders of magnitude compared to proton beams. 

Exclusive vector meson photoproduction, where a vector meson is produced in an event with no other final state particles, is of particular interest. In Pb--Pb collisions this measurement sheds light on nuclear gluon shadowing in the low Bjorken-x region, where it is poorly known. On the other hand p--Pb collisions allow a detailed study of the gluon distribution in the proton and provides a powerful tool to search for gluon saturation. In this talk, results on charmonium (\Jpsi~and \Ppsi) photoproduction measured by the ALICE Collaboration at the LHC are presented \cite{Ref:UPC1} \cite{Ref:UPC2} \cite{Ref:UPC3}. 

\section{ALICE experiment}
The ALICE experiment \cite{Ref:ALICE} consists of a central barrel placed in a large solenoid magnet ($B$ = 0.5 T), covering the pseudorapidity region $|\eta| <$ 0.9, and a muon spectrometer at forward rapidity, covering the range $-4.0 < \eta < -2.5$. 

The analysis at mid-rapidity makes use of the barrel detectors (SPD, TPC and TOF), the VZERO counters and the ZDC calorimeters. The Silicon Pixel Detector (SPD) makes up the two innermost layers of the ALICE Inner Tracking System (ITS) It is a fine granularity detector, having about 107 pixels, and can be used for triggering purposes. The other four layers of the ITS are used in this analysis for particle tracking. The Time Projection Chamber (TPC) is used for tracking and for particle identification and has an acceptance covering the pseudorapidity region $ |\eta| < $ 0.9. Beyond the TPC, the Time-of-Flight detector (TOF) is a large cylindrical barrel of Multi-gap Resistive Plate Chambers. Its pseudorapidity coverage matches that of the TPC. Used in combination with the tracking system, the TOF detector can be used for charged particle identification up to about 2.5 GeV/c (pions and kaons) and 4 GeV/c (protons). In addition the TOF detector can be used for triggering purpose too.
The VZERO counters consist of two arrays of 32 scintillator tiles and finally, two sets of hadronic Zero Degree Calorimeters (ZDC) detect neutrons emitted in the very forward region such as neutrons produced by electromagnetic dissociation.

The forward rapidity analysis is based on the muon spectrometer. The spectrometer consists of a ten interaction length thick absorber filtering the muons, in front of five tracking stations containing two planes of cathode pad multi-wire proportional chambers each, with the third station placed inside a dipole magnet. The forward muon spectrometer includes a triggering system, used to select muon candidates with a transverse momentum larger than a given programmable threshold.

The experimental challenge for such measurements consists in having dedicated UPC triggers that are often orthogonal to the general trigger strategy of the experiments. The forward trigger is based on the muon spectrometer trigger system which requires the presence of an unlike sign dimuon candidate where both tracks have \pt~$>$ 0.5 GeV/c and the VZERO trigger set for at least one cell in VZERO-C (muon arm side) and empty for VZERO-A (opposite side).

The trigger for central rapidity is based on at least two and less than six hits in TOF, with two of them back-to-back in azimuth, requirement on SPD fired chips specific for each collision system and empty VZERO detectors on both sides.

The luminosity collected by UPC triggers during the 2011 Pb-Pb run was 55 $\rm nb^{-1}$ in the forward sample and 23 $\rm nb^{-1}$ in the central rapidity sample. In the LHC p-Pb run (both directions) we collected 9.4 $\rm nb^{-1}$ for the forward trigger.

\section{\Jpsi~photoproduction in Pb--Pb collisions}

The exclusive photoproduction can be either coherent, where the photon couples coherently to almost all the nucleons, or incoherent, where the photon couples to a single nucleon. Coherent production is characterized by low transverse momentum of the produced vector mesons (\pt $\approx$ 60 MeV/c), and the nucleus normally does not break up in the \Jpsi~production. Incoherent production, corresponding to quasi-elastic scattering off a single nucleon, is characterized by a somewhat higher transverse momentum (\pt $\approx$  500 MeV/$c$). In this case the nucleus interacting with the photon breaks up, but, apart from single nucleons or nuclear fragments in the very forward region, no other particles are produced.

\begin{figure}[htb]
\centering
\includegraphics[width=0.49\textwidth]{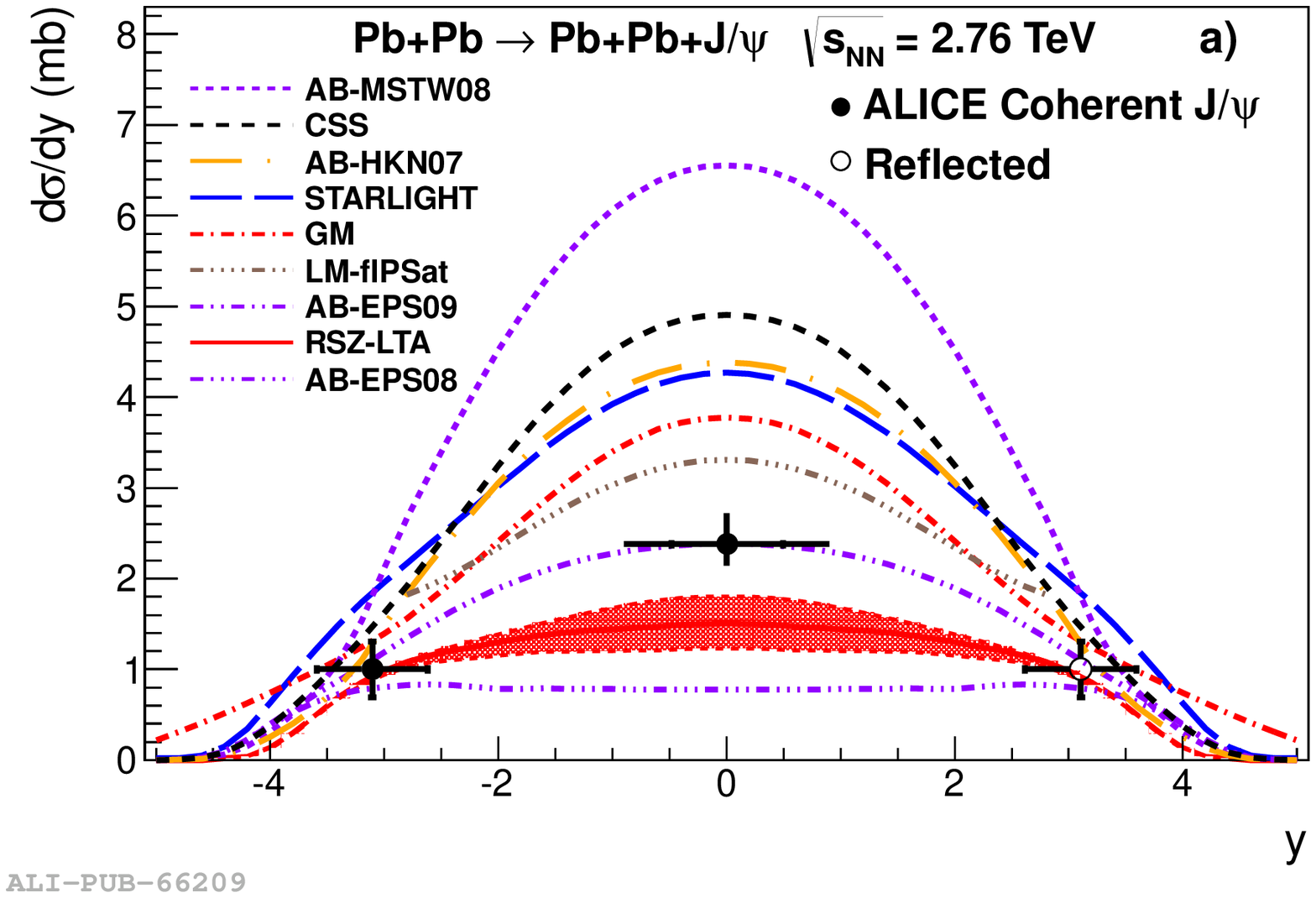}
\includegraphics[width=0.49\textwidth]{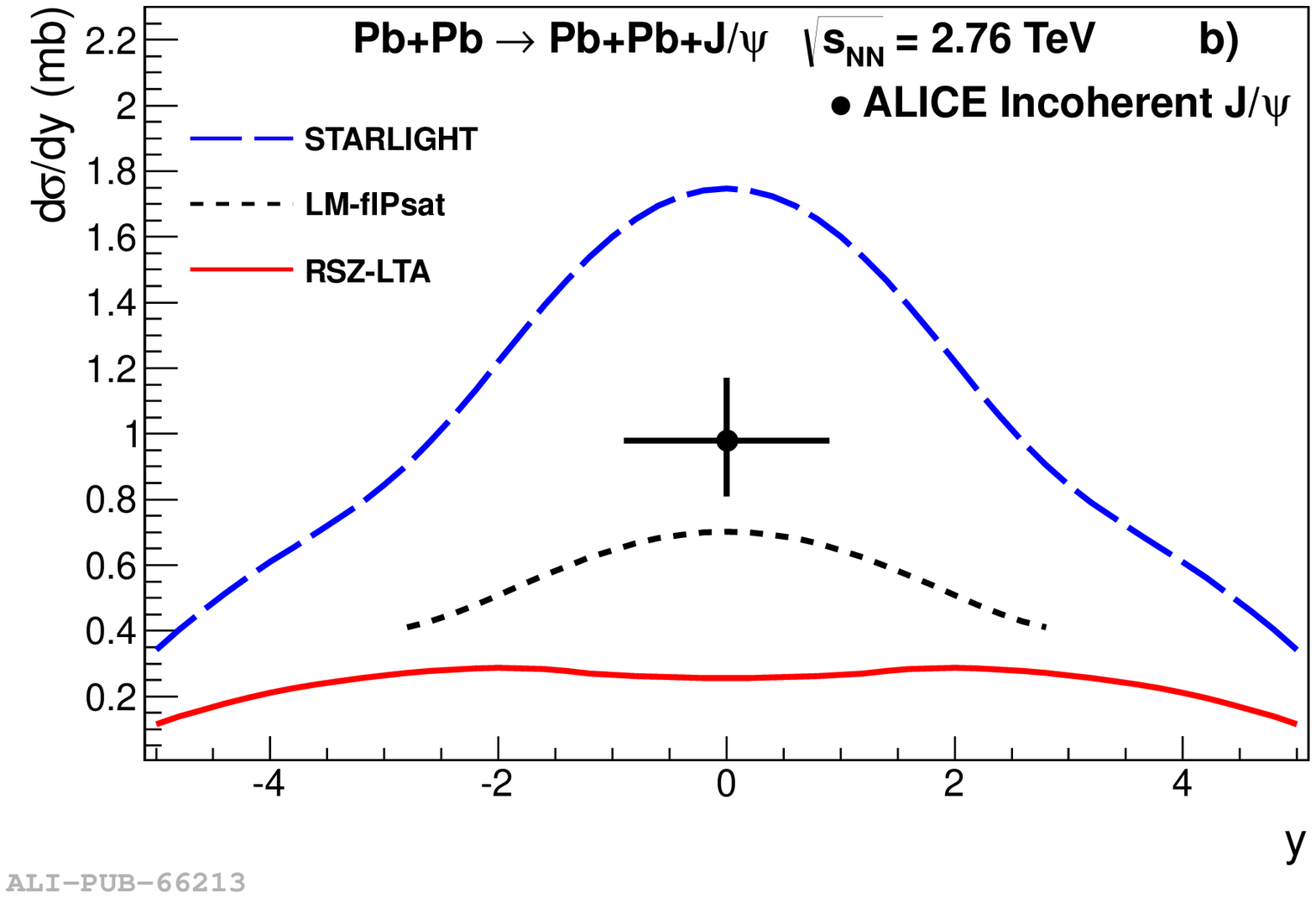}
\caption{Measured differential cross section of \Jpsi~photoproduction
in ultra-peripheral Pb--Pb collisions at $\sqrt{s_{\rm NN}} = 2.76$~TeV for coherent (left) and incoherent (right) events. The error is the quadratic sum of the statistical and systematic errors. The theoretical calculations described in the text are also shown}
\label{fig:JPsi}
\end{figure}

The \pt~distributions for both coherent-enhanced and incoherent-enhanced data samples are well described by fitting together the expected signals and
backgrounds using templates built from MC and data. This was done to extract the measured yield of coherent and incoherent \Jpsi~candidates. The measured cross sections are  $\rm {d}\sigma^{coh}_{J/\psi}/\rm{d} y = 1.00 \pm 0.18$ (stat)~$^{+0.24}_{-0.26}$ (syst) mb at -3.6 $<$ y $<$ -2.6 and $\rm {d}\sigma^{coh}_{J/\psi}/\rm{d} y = 2.38~^{+0.34}_{-0.24} $(stat+syst) mb at -0.9 $<$ $y$ $<$ 0.9. These results were compared to the available model calculations as described in [2, 3]. The best agreement is found with models that include nuclear gluon shadowing, consistent with EPS09 (see Fig.\ref{fig:JPsi} left). Furthermore, the incoherent \Jpsi~cross section was also recently published as $\rm {d}\sigma^{incoh}_{J/\psi}/\rm{d} y = 0.98~^{+0.19}_{-0.17} $(stat+syst) mb at -0.9 $<$ $y$ $<$ 0.9 and provides additional constraints for model calculations (see Fig.\ref{fig:JPsi} right).

\section{Coherent \Ppsi~ photoproduction in Pb--Pb collisions} 

Photoproduced \Ppsi~mesons can be studied in the dilepton decay channel or in the decay channel  $\rm \psi(2s) \rightarrow J/\psi + \pi^{+} \pi^{-}$ followed by $\rm J/\psi \rightarrow l^{+} l^{-}$, which has a much higher branching ratio. These events are characterized by two hard tracks with \pt$>$1 GeV/c from the decay of the \Jpsi~and two soft tracks with \pt$<$0.4 GeV/$c$ from the two pions. The triggered events were required to have exactly 2 or 4 reconstructed tracks and a reconstructed primary vertex. The dileptons can be separated into electrons or muons using the TPC d$E$/d$x$. To select coherently produced \Ppsi~a requirement \pt$<$0.15 (0.30) GeV/$c$ was applied for the decay channels containing muons (electrons). The yield in the direct dilepton channel ($\rm \psi(2s) \rightarrow\mu^{+} \mu^{-}$ and $\rm \psi(2s) \rightarrow e^{+} e^{-}$ combined) was obtained by fitting the invariant mass distribution to the sum of a Crystal Ball function for the signal and an exponential for the background. 

The 4 track samples have a very high purity and the yield is obtained by counting the number of events within selected intervals in phase space determined from Monte Carlo simulations. The invariant mass of the dileptons is required to be in the interval 3.0$<M_{Inv}<$3.2 (2.6$<M_{Inv}<$3.2) GeV/$c^2$ for muon (electron) channel to select pairs coming from a \Jpsi~decay. The $\mu^{+} \mu^{-} + \pi^{+} \pi^{-}$ ($e^{+} e^{-} + \pi^{+} \pi^{-}$) final state is required to have an invariant mass 3.6$<M_{Inv}<$3.8 (3.1$<M_{Inv}<$3.8) GeV/$c^2$ to count as a \Ppsi. This gives a total of 17 (11) events for the $\mu^{+} \mu^{-} + \pi^{+} \pi^{-}$ ($e^{+} e^{-} + \pi^{+} \pi^{-}$) decay channel, with a like-sign background of 1 (0) event. The extracted yield is corrected for acceptance and effciency using STARLIGHT \cite{Ref:STARLIGHT} events processed through the detector response simulation and reconstructed using the same selection as for real data. 

\begin{figure}[htb]
\centering
\includegraphics[width=0.49\textwidth]{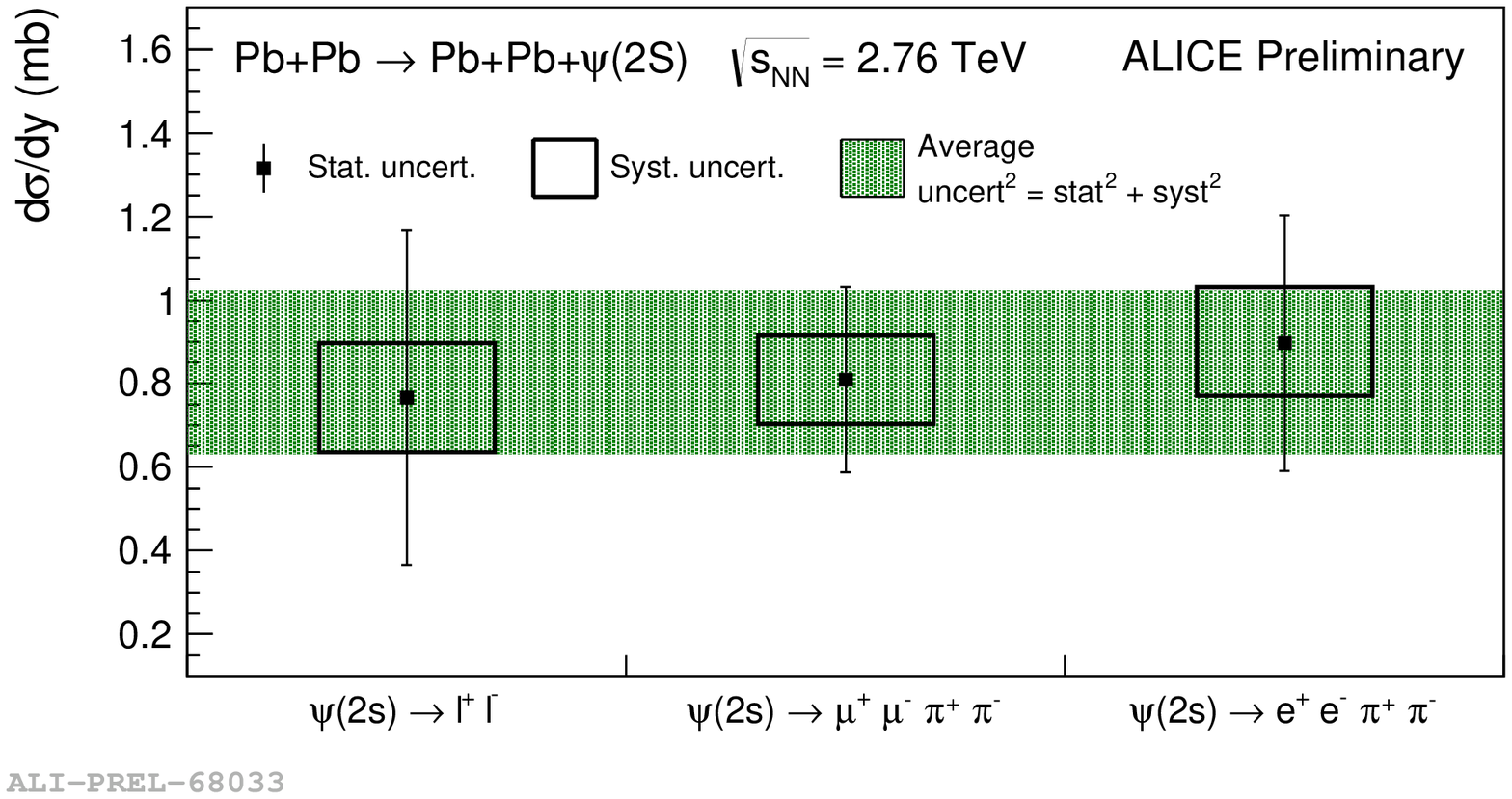}
\includegraphics[width=0.49\textwidth]{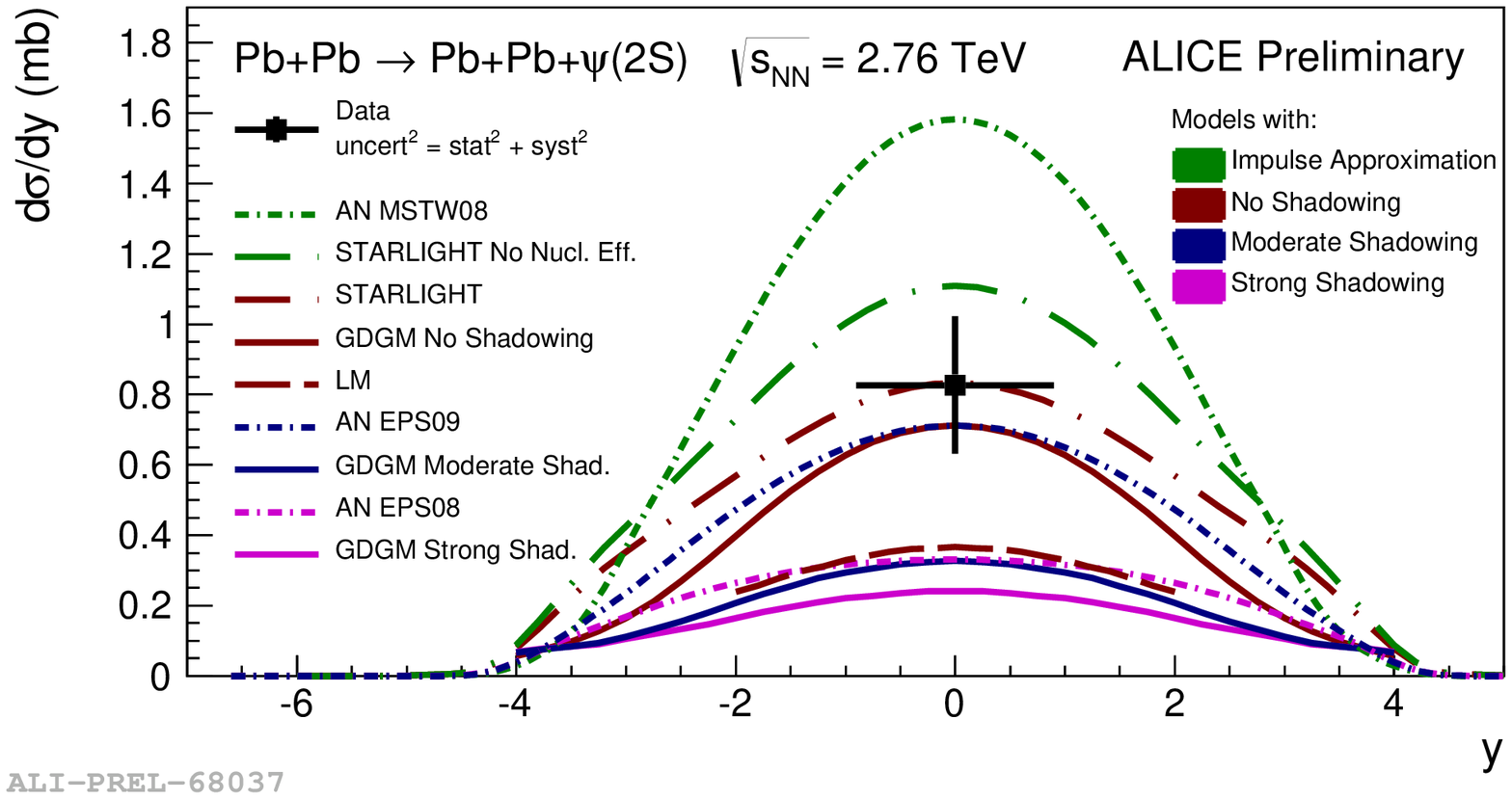}
\caption{Measured differential cross section of \Ppsi~photo-production in Pb-Pb ultra-peripheral collisions at $\sqrt{s_{\rm NN}} = 2.76$~TeV at -0.9 $<$ y $<$ 0.9, in three different channels(left) and in comparison with models (right).}
\label{fig:Psi2s}
\end{figure}

The cross section is obtained from a weighted mean of the three decay channels (see Fig.~\ref{fig:Psi2s} left), giving $\rm {d}\sigma^{coh}_{\psi(2s)}/\rm{d} y = 0.827~^{+0.19}_{-0.19}$ (stat+syst) mb. This is compared to model predictions \cite{Ref:AN} \cite{Ref:GDGM} \cite{Ref:LM} in Fig~\ref{fig:Psi2s} right. The \Ppsi~is a hard probe, the scale being set by its mass, and is therefore expected to be sensitive to the nuclear gluon distribution in the same way as the \Jpsi. The experimental error is, however, larger for the \Ppsi~than for the \Jpsi~measurement, because of the limited statistics. It is also important to note that the underlying $\gamma$ + p $\rightarrow$ V + p cross section has a considerably larger uncertainty for \Ppsi~than for \Jpsi. For \Jpsi~a wealth of $\gamma$ + p $\rightarrow$ \Jpsi~+ p cross section data has been obtained by ZEUS \cite{Ref:ZEUS} and H1 \cite{Ref:H11} \cite{Ref:H12}, while the process $\gamma$ + p $\rightarrow$ \Ppsi~+ p was measured by H1 at four different energies only. This makes the theoretical cross section constraints on the experimental data much weaker. This is illustrated by the large discrepancy between the AN-MSTW08 and STARLIGHT predictions without nuclear effects in Fig~\ref{fig:Psi2s} right. One can nevertheless conclude that the measured cross section disfavours models with no nuclear effects and those with strong gluon shadowing.

\begin{figure}[htb]
\centering
\includegraphics[width=0.49\textwidth]{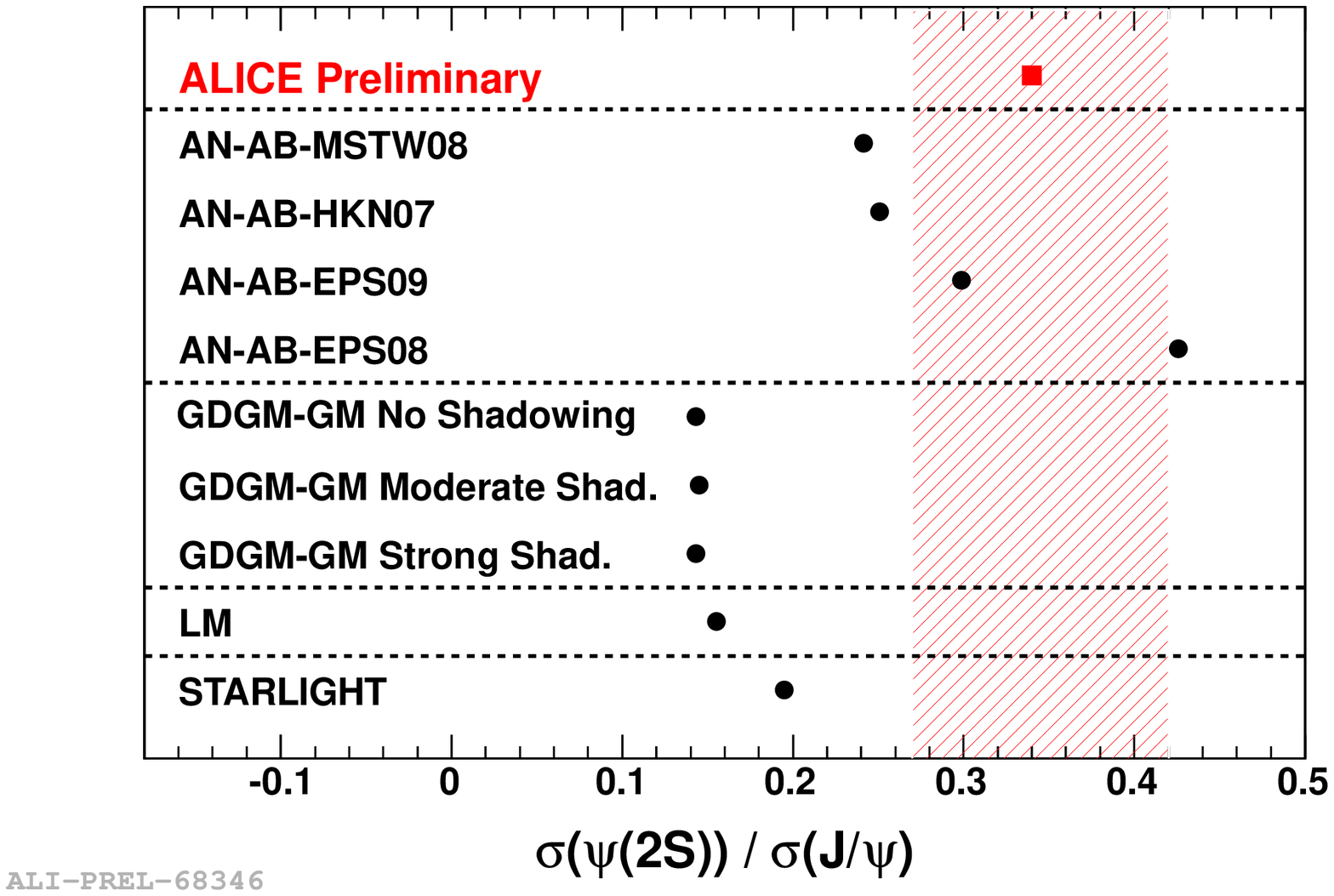}
\includegraphics[width=0.49\textwidth]{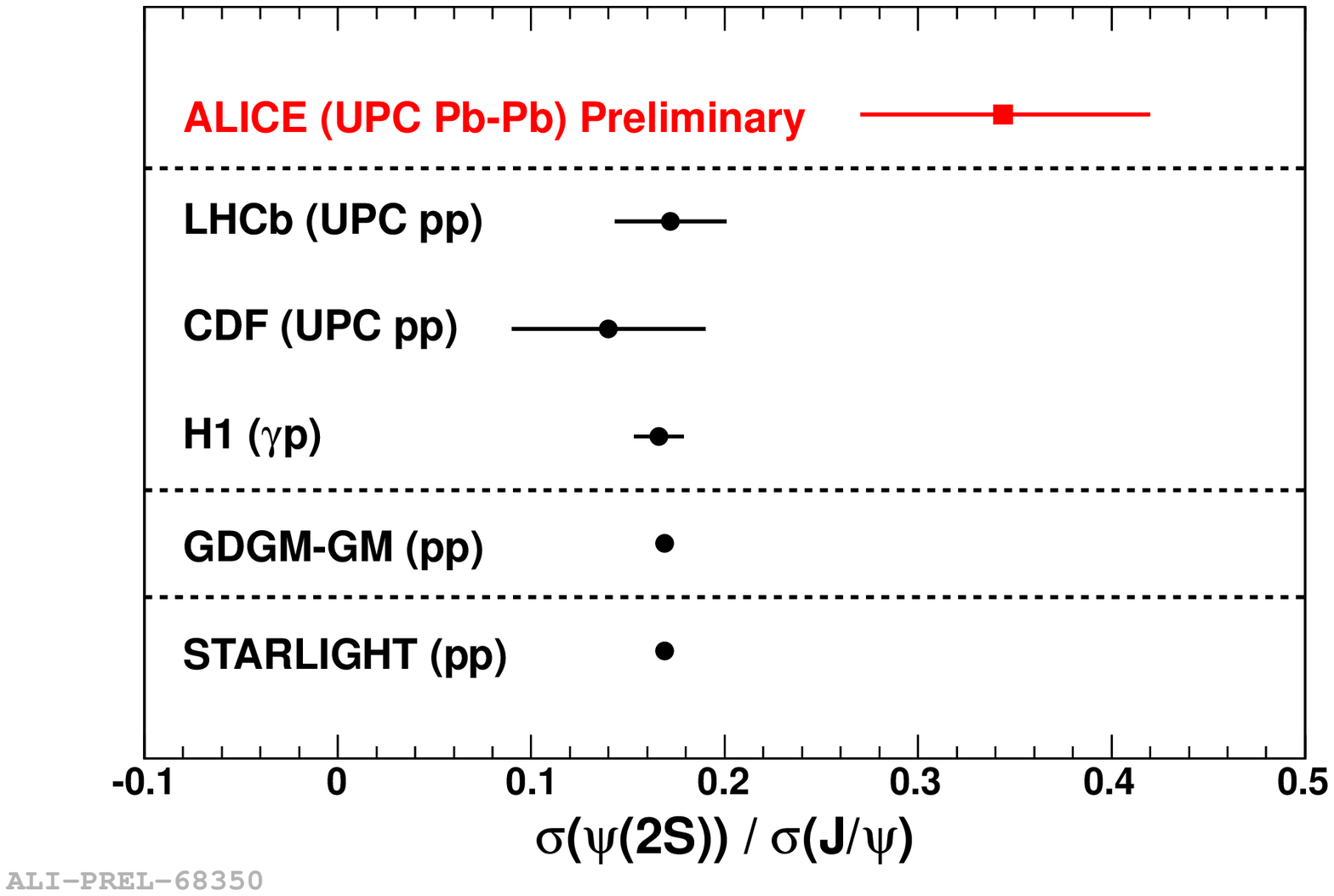}
\caption{Ratio of the \Ppsi~to \Jpsi~cross section, for pp and $\gamma$p interactions compared to theoretical predictions(left), and in Pb-Pb collisions (right).}
\label{fig:Ratio}
\end{figure}

In order to compare the coherent \Ppsi~cross section with the previously measured \Jpsi~cross section, we give the ratio of the two. Many of the systematic uncertainties of these measurements are correlated and cancel out in the ratio. The final ratio gives a value of $\sigma^{coh}_{\psi(2s)}/\sigma^{coh}_{J/\psi} = 0.344~^{+0.076}_{-0.074}$. Fig.\ref{fig:Ratio} (left) shows the \Ppsi~to \Jpsi~cross section ratio measured in pp collisions by CDF and LHCb. Both STARLIGHT and the model by Gay Ducati, Griep, and Machado predict the data correctly. The figure also shows the ratio measured by H1 in $\gamma$-p collisions and the Pb--Pb ratio obtained with the present analysis. The H1 result is compatible with the pp measurements, while the ALICE point is a factor 2 larger, although still with sizable errors. This difference may indicate the nuclear effects and/or the gluon shadowing modify the \Jpsi~and the \Ppsi~production in a different way. Fig.\ref{fig:Ratio} (left) shows the comparison of the \Ppsi~to \Jpsi~cross section ratio between measurements and predictions in Pb--Pb collisions \cite{Ref:ZH} \cite{Ref:GM}. The models predict a Pb--Pb ratio a factor 2 smaller than the one measured by ALICE. It is worth noting models that reproduced the pp ratio correctly, fail in describing the Pb--Pb ratio. It is surprising the AN model, although it assumes a \Ppsi~wave function identical to the \Jpsi~one, describes this ratio in a satisfactory way.

\section{\Jpsi~photoproduction in p--Pb collisions}

Photons produced by a high energy nucleus colliding with a proton, offer the possibility to shed light on the proton structure functions. The lead nucleus acts as a photon emitter (enhanced flux by factor Z2  compared with the proton photon flux). At LHC the Bjorken-$x$ goes down to $10^{-5}$ reaching for the first time the 1 TeV scale for the $\gamma$p centre-of-mass energy ($\rm W_{\gamma p}$). 

The LHC beams were reversed in order to measure both forward and backward rapidity. Thus, \Jpsi~ measured in the muon spectromenter at forward rapidity via $\rm J/\psi \rightarrow\mu^{+} \mu^{-}$ decay channel are reconstructed in the 2.5 $<$ y $<$ 4.0 (p--Pb) and 3.6 $<$ y $<$ 2.6 (Pb--p) rapidity intervals, where y is measured in the laboratory frame with respect to the proton beam direction. The $W_{\gamma p}$ is determined by the \Jpsi~rapidity: $W^{2}_{\gamma p} = 2 E_{p} M_{J/\psi} e^{-y}$, where $M_{J/\psi}$ is the \Jpsi~mass, $y$ is the \Jpsi~rapidity and $E_{p}$ is the proton energy ($E_{p}$ =4 TeV in the lab frame), while the Bjorken-$x$ is given by $x=(M_{J/\psi}/W_{\gamma p})^{2}$. We study 21$< W_{\gamma p} <$45 GeV for y $>$ 0 and 577$< W_{\gamma p} <$952 GeV for $y$ $<$ 0, thereby exceeding the $\gamma$p range of HERA.

\begin{figure}[htb]
\centering
\includegraphics[width=0.7\textwidth]{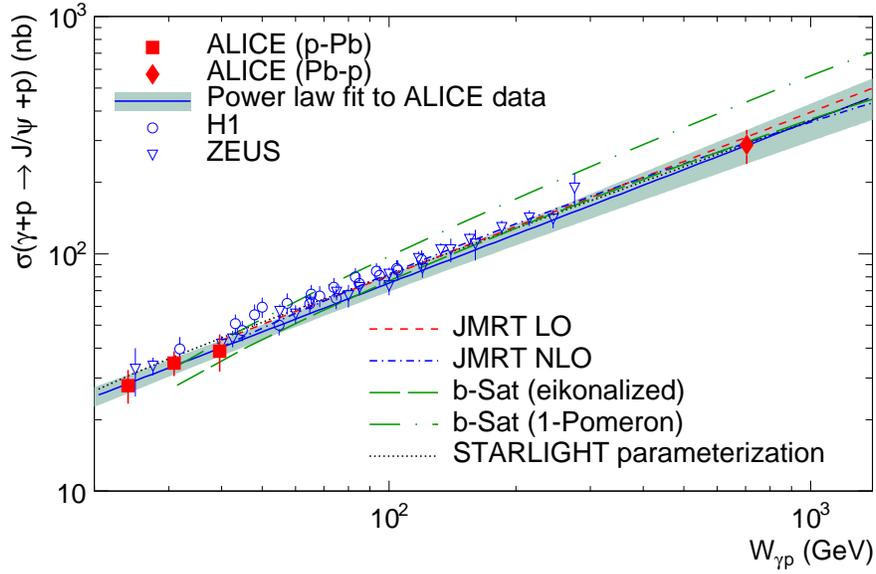}
\caption{Exclusive \Jpsi~photoproduction cross section off protons measured by ALICE and compared to HERA data. Comparisons to STARLIGHT, JMRT and the b-Sat models are shown. The power law fit to ALICE data is also shown.}
\label{fig:pPb}
\end{figure}

Fig.\ref{fig:pPb} shows the ALICE measurements for $\sigma(W_{\gamma p})$. Comparisons to previous measurements and to different theoretical models are also shown. As mentioned earlier, $\sigma(W_{\gamma p})$ is proportional to the square of the gluon PDF of the proton \cite{Ref:Ryskin}. For HERA energies, the gluon distribution at low Bjorken-$x$ is well described by a power law in x, which implies the cross section $\sigma(W_{\gamma p})$ will also follow a power law \cite{Ref:xDep}. A deviation from such a trend in the measured cross section as $x$ decreases, or equivalently, as $W_{\gamma p}$ increases, could indicate a change in the evolution of the gluon density function, as expected at the onset of saturation.

Both ZEUS and H1 fitted their data using a power law  $\sigma \approx W_{\gamma p}^{\delta}$, obtaining $\delta$ = 0.69 $\pm$ 0.02 (stat) $\pm$ 0.03 (syst), and $\delta$ = 0.67 $\pm$ 0.03 (stat + syst), respectively. Due to the large HERA statistics, a simultaneous fit of H1, ZEUS, ALICE low energy points data gives power-law fit parameters almost identical to those obtained from HERA alone. A fit to ALICE data alone gives $\delta$ = 0.68 $\pm$ 0.06 (stat. + syst.). Only uncorrelated systematic errors were considered here. Thus, no deviation from a power law is observed up to about 700 GeV. A natural explanation is that no change in the behaviour of the gluon PDF in the proton is observed between HERA and LHC energies.

\section{Conclusions}

The ALICE Collaboration published the first LHC papers on exclusive \Jpsi~photoproduction in Pb-Pb UPC. Using the Pb-Pb data the coherent and incoherent photoproduction cross sections have been measured. The coherent \Jpsi~cross section is found to be in good agreement with the model which incorporates the nuclear gluon shadowing according to the EPS09 parameterization (AB-EPS09).

The preliminary measurements from ALICE for the coherent \Ppsi~photoproduction in ulta-peripheral Pb--Pb collisions at LHC are described. The result disfavour models considering all nucleons contributing to the scattering (impulse approximation) and those implementing strong shadowing, such as the EPS08 parametrization. Most of the models underpredict \Ppsi~to \Jpsi~cross section ratio by a factor 2. We believe the modelling of the \Ppsi~production in ultra-peripheral collisions requires further efforts.

The ALICE Collaboration also published the first measurement of exclusive \Jpsi~photoproduction off protons in p--Pb collisions at the LHC. Our data are compatible with a power law dependence of $\sigma(W_{\gamma p})$ up to about 700 GeV in $W_{\gamma p}$, corresponding to $x \approx 2\times10^{-5}$. A natural explanation is that no change in the behaviour of the gluon PDF in the proton is observed between HERA and LHC energies.


\begin{thebibliography}{99}

\bibitem{Ref:Baltz}{A. J. Baltz et al., Phys. Rep. 458 (2008) 1.}
\bibitem{Ref:UPC1}{B. Abelev et al. [ALICE Collaboration] Phys. Lett. B 718 (2013) 1273.}
\bibitem{Ref:UPC2}{E. Abbas et al. [ALICE Collaboration], Eur. Phys. J. C 73 (2013) 2617.}
\bibitem{Ref:UPC3}{B. B. Abelev et al. [ALICE Collaboration], arXiv:1406.7819 [nucl-ex].}
\bibitem{Ref:ALICE}{K. Aamodt et al. [ALICE Collaboration], JINST 3 (2008) S08002.}
\bibitem{Ref:STARLIGHT}{S. R. Klein and J. Nystrand, Phys. Rev. C 60, 014903 (1999); A. J. Baltz, Y. Gorbunov, S. R. Klein and J. Nystrand, Phys. Rev. C 80 (2009) 044902; http://starlight.hepforge.org}
\bibitem{Ref:AN}{A. Adeluyi and T. Nguyen, Phys. Rev. C 87 (2013) 027901.}
\bibitem{Ref:GDGM}{M. B. G. Ducati, M. T. Griep and M. V. T. Machado, Phys. Rev. C 88 (2013) 014910.}
\bibitem{Ref:LM}{T. Lappi and H. Mantysaari, Phys. Rev. C 87 (2013) 032201, arXiv:1301.4095 [nucl-th].}
\bibitem{Ref:ZEUS}{S. Chekanov et al. [ZEUSS Collaboration] Eur. Phys. J. C24 (2002) 345.}
\bibitem{Ref:H11}{C. Alexa et al. [H1 Collaboration] Eur. Phys. J C 73 (2013) 2466}
\bibitem{Ref:H12}{C. Adolff et al. [H1 Collaboration] Phys. Lett. B541 (2002) 251.}
\bibitem{Ref:ZH}{L. Frankfurt, M. Strikman, M. Zhalov, Phys. Lett. B 537 (2002) 51; V. Rebyakova, M. Strikman, M. Zhalov, Phys. Lett. B 710 (2012) 647.}
\bibitem{Ref:GM}{V. P. Goncalves and M. V. T. Machado, Phys. Rev. C 84 (2011) 011902.}
\bibitem{Ref:Ryskin}{M. G. Ryskin, Z. Phys. C 57, 89 (1993).}
\bibitem{Ref:xDep}{F. D. Aaron et al. [H1 and ZEUS Collaboration], JHEP 1001, 109 (2010).}


\end{thebibliography}
\end{document}